\documentclass[a4paper,12pt]{article}
\usepackage{amssymb,amsmath,mathrsfs}
\usepackage[usenames,dvipsnames]{xcolor}
\usepackage{hyperref, comment}
\usepackage{tensor}
\usepackage{graphicx}
\usepackage{enumitem}
\usepackage[left=1.2in,top=1in,right=1.2in,bottom=1in,headheight=0.8in,foot=0.5in]{geometry}
\usepackage[nodisplayskipstretch]{setspace} 
\usepackage[width=\textwidth]{caption}
\usepackage{pifont}% http://ctan.org/pkg/pifont

\usepackage[textsize=tiny]{todonotes}

\usepackage{mathtools}

\newcommand{\nocontentsline}[3]{}
\newcommand{\tocless}[2]{\bgroup\let\addcontentsline=\nocontentsline#1{#2}\egroup}
\usepackage[indentafter]{titlesec}
\titleformat{name=\section}{}{\thetitle.}{0.8em}{\centering\scshape}
\titleformat{name=\subsection}[runin]{}{\thetitle.}{0.5em}{\bfseries}[.]
\titleformat{name=\subsubsection}[runin]{}{\thetitle.}{0.5em}{\itshape}[.]
\titleformat{name=\paragraph,numberless}[runin]{}{}{0em}{}[.]
\titlespacing{\paragraph}{0em}{0em}{0.5em}
\titleformat{name=\subparagraph,numberless}[runin]{}{}{0em}{}[.]
\titlespacing{\subparagraph}{0em}{0em}{0.5em}
\hypersetup{
        breaklinks=true,
	colorlinks=true,         
	linkcolor=Blue,          
	citecolor=Blue,
	urlcolor=Blue            
 }

\usepackage[style=numeric,maxcitenames=2,uniquename=true,backend=biber,uniquelist=true,sorting=none,maxbibnames=4]{biblatex}
\DeclareNameAlias{sortname}{last-first}

\AtEveryBibitem{\clearfield{month}}

\renewbibmacro{in:}{}

\usepackage{float}

\title{Navigating permanent underdetermination in dark energy and inflationary cosmology}

\author{William J.~Wolf\footnote{Faculty of Philosophy, University of Oxford, UK. william.wolf@stx.ox.ac.uk}~\footnote{Astrophysics, University of Oxford, UK.}~ \& James Read\footnote{Faculty of Philosophy, University of Oxford, UK. james.read@philosophy.ox.ac.uk}}

\date{}

\makeatletter
\let\uppercasenonmath\@gobble

\begin{document}
\setstretch{1.0}
\maketitle

\begin{abstract}
We identify troubling cases of so-called `permanent underdetermination' in both dark energy and inflationary cosmology. We bring to bear (a) a taxonomy of possible responses to underdetermination, and (b) an understanding of both dark energy and inflationary cosmology from an effective field theory point of view. We argue that, under certain conditions, there are viable responses which can arguably alleviate at least some of the concerns about underdetermination in the dark energy  and inflationary sectors. However, the epistemic threat of permanent underdetermination remains a significant challenge.
\end{abstract}
\tableofcontents
\setstretch{1.2}

\newpage

\section{Introduction}

The standard `$\Lambda$CDM + inflation' model of modern cosmology is remarkably successful in accurately describing the evolution of the universe from mere fractions of a second after its birth until the present day \cite{Planck:2018vyg}. Notwithstanding a few anomalies, all the available evidence indicates that this model offers an excellent description of reality. Yet, there remains a persistent sense of dissatisfaction due to the glaring absence of adequate explanations for much of the model's structure, which stems from the fact that it is largely phenomenological in nature. 
% The basic ingredients of the model include:
% \begin{description}
%     \item[Friedmann–Lema\^itre–Robertson–Walker (FLRW) metric:] The universe is described on large scales by the FLRW geometry, which is characterized by its homogeneity and isotropy. Deviations from homogeneity and isotropy are treated as small perturbations. 
%     \item[Inflation:] An early period of accelerated expansion that smoothed and flattened the universe, and produced tiny density perturbations that seeded future large-scale structure, driven by a field called the `inflaton'.
%     \item[Baryonic matter and radiation:] Matter-energy content represented by the familiar standard model of particle physics.
%     \item[Dark matter:] A non-baryonic `dark' matter that is crucial for accounting for empirical observations of galaxy rotation curves, the matter power spectrum, gravitational lensing, etc.
%     \item[Dark energy:] A late period of accelerated expansion that the universe is only just entering driven by a form of `dark' energy.
% \end{description}
The reasons for dissatisfaction are obvious. The only component of the model over which we have any kind of firm epistemic control are the fields in the standard model of particle physics, and these represent only a tiny fraction of the universe's energy budget at $\sim 5\%$ (compared with $\sim 25\%$ for dark matter and $\sim 70\%$ for dark energy). In the words of \textcite[p.~340]{Peebles2020}, the model consists of placeholders that represent the ``simplest ideas that would allow a fit to the observations'': `$\Lambda$' refers to a cosmological constant, `CDM' refers to cold dark matter, and `inflation' refers to a dynamical scalar field in the very early universe.

One of the goals of modern cosmology is to determine the `underlying physical theory' \cite[p.~3]{DiValentino:2021izs} behind this effective description of the universe. However, recent developments in cosmology indicate that this goal---already recognized as exceptionally challenging---might be even more daunting than cosmologists had expected. In particular, \textcite{FWR} consider seriously the possibility that cosmological observations will permanently underdetermine the microphysical models underlying the phenomena behind inflation, dark matter, and dark energy due to the limited amount and kind of empirical information that can be extracted from them. The variety of model-building constructs that exist within current cosmology are \textit{very} broad for all of these three exotic energy components; here, we will zoom in on this claim with respect to certain classes of inflation and dark energy models, illustrating in detail how the simplest classes of inflation and dark energy models (i.e.\ canonical, single scalar field models) are permanently underdetermined with respect to the primary cosmological observables in their respective contexts. We then investigate and apply a philosophical taxonomy of possible responses (that was previously developed in the context of strong underdetermination) to these instances of permanent underdetermination, arguing that some of these theories' effective field theory (EFT) formulations map onto these philosophical responses and finding that under some circumstances the underdetermination within these restricted classes of theories can arguably be defused (in a sense which we'll explain further below).

The structure of the paper is as follows. \S\ref{Sec:cosmology} reviews recent developments in inflationary and dark energy model building, and how cosmologists map between these theories and cosmological observables. \S\ref{Sec:underdetermination} argues that model building in both dark energy and inflation reflect instances of what \textcite{Pitts2010-PITPUF} has called `permanent underdetermination', in the sense that there will always be distinct microphysical theories that attribute fundamentally different structures to nature, but which give empirical predictions that are \textit{arbitrarily close} to each other; meaning that their underdetermination can never be broken empirically. §\ref{sec:eft} introduces effective field theories (EFTs), as these will feature prominently in analyzing the strategies that have been pursued in response to permanent underdetermination.
\S\ref{Sec:responses_in_cosmo} explores and assesses applications of the discrimination, overarching, and common core approaches (in the terminology of \textcite{LeBihan2018-LEBDAO}) in response to permanent underdetermination in dark energy and inflationary cosmology, and argues that there are some viable strategies that can break the underdetermination.

\section{State of play in modern cosmology}\label{Sec:cosmology}

The simplest versions of inflation and dark energy theories are both given by a single, canonical scalar field on an FLRW metric:
\begin{equation}
S=\int d^4 x \sqrt{-g}\left[\frac{1}{2} M_{\mathrm{pl}}^2 R-\frac{1}{2} g^{\mu \nu} \partial_\mu \varphi \partial_\nu \varphi-V(\varphi)\right],
\end{equation}
where $g$ is the metric, $R$ is the Ricci scalar, $M_{\mathrm{pl}}$ is the Planck mass, $\varphi$ is the scalar field, $V(\varphi)$ is the scalar field potential. When modeling the early universe, this theory is referred to as `inflation'
%and the scalar field is taken to dominate the mass-energy budget of the universe. 
and when modeling dark energy in the late time universe, this theory is referred to as `quintessence'.\footnote{While the action above is written with a minimal coupling between the scalar field and the Ricci scalar, in the inflation paradigm it is common to also consider non-minimal couplings between the scalar field and gravity as there are plausible arguments that they are to be expected at these energies \parencite{Martin:2013tda}. Such non-minimal couplings can also be considered in quintessence, but since this is less common than in inflation we will follow the main physics literature here and confine ourselves to minimally coupled quintessence models (see \textcite{Tsujikawa:2013fta} for a comprehensive review).}

\subsection{Inflation}

Inflation initially gained traction due to its ability to offer satisfying explanations for various fine-tuning problems within the Hot Big Bang model \cite{Guth:1980zm, Starobinsky:1980te},\footnote{The very brief characterization here glosses over some details. See e.g.\ \cite{McCoy2015-MCCDIS, McCoy2019-MCCEJA, WolfDuerr, Smeenk2005-SMEQVE} for further discussion on the nature and severity of these fine-tuning problems, inflation's achievements in explanatory power and predictive novelty, and various other theoretical motivations at play in the context of inflation's development.} such as its ability to answer the question, `why is the universe so precisely flat and homogeneous?' Inflation offers a compelling dynamical resolution to those problems by introducing a scalar field $\varphi$ with a potential $V(\varphi)$ that dominates the matter-energy content of the universe at early times.\footnote{While inflation is by far the most popular framework for modeling the early universe, it does have a number of unresolved issues that have led critics to pursue alternatives. This will not be discussed here, but see \cite{Ijjas:2013vea, Guth:2013sya, Dawid:2023mmv, Wolf:2022yvd} for further discussion.} While many different functional forms of the potential have been considered, all giving distinct microphysical models of inflation (e.g.\ the interaction responsible for inflation could be given by massive fields, exponentials, axions, Nambu--Goldstone bosons, the Higgs or Higgs-like fields, etc.), as long as the potential is sufficiently flat it can alleviate these fine-tuning concerns. 
A crucial quantity here is the so-called `equation of state', defined by $w \equiv p/\rho$, which is the ratio of pressure $p$ and energy density $\rho$ of a perfect fluid. The forms of the equations of state of the various energy density components within the universe will determine the dynamical trajectory of spacetime through the Friedmann equations. When the universe is dominated by a scalar field with a flat potential, 
this generates an equation of state $w(a)\simeq-1$, which effectively acts as a repulsive form of gravity and causes the universe's scale factor $a$ to expand quasi-exponentially in time, $a \simeq e^{Ht}$.
This famously solves both the flatness and horizon problems (see \textcite[Ch.~4]{Baumann:2022mni} for details).

Yet, where inflation truly shines is its account of cosmic structure. Inflation generically predicts that quantum fluctuations in the scalar field should produce slight deviations from uniformity, and that these scalar perturbations should be approximately adiabatic, Gaussian, and scale-invariant. Primordial perturbations matching this description have been observed, and it is these perturbations that source the large-scale structure in the late-time universe \cite{Planck:2018vyg, Planck:2018jri}.

In addition to these scalar perturbations, inflation is also expected to produce tensor perturbations, 
with their amplitudes and power spectra being denoted, $A_s$ and $A_t$, and $\mathcal{P}_s$ and $\mathcal{P}_t$, respectively. 
As mentioned above, the amplitude and power spectra of the scalar fluctuations have been measured; however, the tensor fluctuations (i.e.\ primordial gravitational waves) still elude detection and are one of the primary targets of ongoing and future cosmological probes. 
Crucially, the dynamics of individual inflationary models generally give predictions for the ratio of the amplitudes of scalar and tensor perturbations, as well as for the scale-dependence of the scalar fluctuations. Thus, inflation is characterized primarily by two observables, the tensor-to-scalar ratio $r$ and the scalar spectral index $n_s$:
\begin{equation}
r = \frac{A_s}{A_t}, \qquad n_s(k)-1=\frac{d \ln \mathcal{P}_s}{d \ln k},
\end{equation}
both of which can be computed directly from an inflation model.

While many models of inflation do map onto distinct regions of the ($r$, $n_s$) parameter space (see \cite[Fig.~8]{Planck:2018jri} for the inflationary `zoo plot' of models) and there was initially the general expectation that inflation should produce an observable $r$ \cite{Boyle:2005ug, Tegmark:2004qd}, as the upper bound on $r$ has been pushed lower and as theorists have further explored the inflationary landscape, these initial expectations have proved to be too naïve. 

To list just a few examples, \textcite{Kallosh:2018zsi} demonstrated how one can cover the entire viable region of ($r$, $n_s$) plane with `$\alpha$-attractor' and `KKLT' models.
\textcite{Stein:2022cpk, Wolf:2024lbf} showed that, within `hilltop' models, higher order terms in the potential, which were often neglected in computing their predictions, in fact can have a significant effect on the end of inflation and can reduce predictions for $r$ arbitrarily while still remaining within the viable $n_s$ region. \textcite{Sousa:2023unz} used machine learning techniques to identify inflationary potentials and found several largely unexplored functional forms with predictions below observational thresholds in the ($r$, $n_s$) plane. All of these constructions can be understood within the simplest version of the inflationary paradigm and do not generate any egregious added complexity or \textit{ad hocness}.
Yet, they are distinctly different microphysical accounts in terms of the fundamental interactions which they take to underlie inflation.
Furthermore, the constructs mentioned here all have the ability to push ($r$, $n_s$) many orders of magnitude below projected experimental sensitivities for next generation CMB probes \parencite{CMB-S4:2016ple} in addition to covering much, if not all, of the remaining viable parameter space. 
%that will remain within experimental reach.
%Furthermore, the constructs mentioned here all have the ability to push ($r$, $n_s$) many orders of magnitude below projected experimental sensitivities for next generation CMB probes \cite{CMB-S4:2016ple}.

\subsection{Dark energy}
The presence of dark energy is inferred primarily through distance measurements \cite{Frieman:2008sn}. That is, cosmological observables such as angular diameter distances or luminosity distances are sensitive to the Hubble rate $H(a)$, which relates the universe's rate of expansion in terms of its scale factor $a$ to its energy density through the Friedmann equation. Until a few decades ago, cosmologists assumed that radiation and matter were the only stress-energy species relevant to the dynamics of the universe. However, if we assume they are the \textit{only} sources of energy density in our cosmological modelling, there are large discrepancies between the cosmological distances observed and those predicted under those modeling assumptions \cite{Durrer:2011gq}. These observations indicate that there is a missing component in the universe's energy density.

In other words, $H(a)$ can be rewritten in the following way to display its sensitive to how various types of energy density scale with respect to the expansion of the universe's scale factor:
\begin{equation}\label{eq:hubble_fit}
H^2(a) = H_0^2 \left[ \Omega_{\mathrm{r}} a^{-4} +  \Omega_{\mathrm{m}} a^{-3} + \Omega_{\mathrm{x}} e^{3 \int_a^1\left(1+w_{\text {x }}\right) \mathrm{d} \ln \mathrm{a}} \right],
\end{equation}
where $\Omega_x$ represents the energy density and $w_x$ represents the equation of state for some unspecified additional component. Taking $w_x \equiv w_{\rm DE} \simeq -1$ and $\Omega_x \equiv \Omega_{\rm DE} \simeq 0.7$ brings the predicted and observed distance measurements into alignment. This indicates that the universe is dominated by a form of `dark' energy that is (approximately) not diluting with the increase of the scale factor; thus entering another period of accelerated, quasi-exponential expansion, in close analogy with the inflationary account of the early universe.

How do we map between the data/observational side and the theory space of dark energy? As the effects of dark energy models are primarily driven by the behavior of their equation of state, physicists have largely adopted a well-known parameterization of the dark energy equation of state known as the Chevallier--Polarski--Linder (CPL) parameterization \cite{Linder:2002et, Chevallier:2000qy}:
\begin{equation}\label{param}
w(a)=w_0+w_a(1-a),
\end{equation}
where $w_0$ is the (inferred) value of the equation of state today and $w_a$ characterizes its temporal variation. This allows us to characterize various dark energy models in terms of the pairs ($w_0$, $w_a$). For example, $\Lambda$ would be given by ($-1$, $0$), while any dynamical model would have $w_a\neq0$. If dark energy is dynamical (i.e.\ not driven by $\Lambda$), the next most simple and obvious way to model it is to adapt the single scalar field machinery of inflation to the dark energy problem, as was most notably done by \textcite{Ratra:1987rm, Caldwell:1997ii}, which is known as `quintessence'. While the observational picture here is still far from settled, recent results from the DESI collaboration \cite{DESI:2024mwx, DESI:2025zgx} have provided the first substantial evidence for deviations from a cosmological constant, favoring a dynamically evolving equation of state with a $4\sigma$ statistical significance.\footnote{These new results have generated much recent discussion and debate in the physics literature. See e.g.\ \cite{Cortes:2024lgw, Wolf:2024eph, Shlivko:2024llw, Dinda:2024ktd, Wolf:2024stt, Ye:2024zpk, Jiang:2024viw, Jiang:2024xnu, RoyChoudhury:2024wri, Wolf:2025jed, DESI:2024kob, Wolf:2025acj, Chakraborty:2025syu} for a representative sample of recent analyses.} At the very least, these results motivate considering a dynamical framework that goes beyond the base cosmological constant scenario, and provide an additional urgency in investigating dynamical proposals such as quintessence.

As with inflation, there was some hope that cosmologists would be able to whittle down substantially, or perhaps pin down precisely, microphysical models of dark energy by their predictions for ($w_0$, $w_a$) \cite{Caldwell:2005tm}. Yet, these hopes have likewise not materialized. More specifically, current constraints highly favour the `thawing' regime of dark energy (meaning dark energy is getting weaker).
`Hilltop' models of quintessence
have dynamical features that enable them to describe the equation of state $w(a)$ as evolving in a slow, approximately linear manner, or in a very rapid, highly non-linear manner, and everything in between \parencite{Wolf:2023uno, Wolf:2024eph, Shlivko:2024llw, Dutta:2008qn}. Consequently, these models can arbitrarily saturate huge swathes of the ($w_0$, $w_a$) parameter space because they can effectively generate a slow dynamical evolution, in which case they approximate the universal behavior of the many familiar models found in \cite{Scherrer:2007pu}, or an arbitrarily rapid dynamical evolution (captured in $w_a$) for any value of the equation of state today $w_0$, in which case they approximate a number of other distinct models with similar features in their potentials.\footnote{There is an additional nuance here as there are a few different interpretations of the $(w_0, w_a)$ parameterization. Often, it is interpreted as a Taylor expansion of $w(a)$ around recent cosmological times and the literature will look to the dark energy model's equation of state at a particular redshift or over a certain redshift range as indicative a particular model's representation in the $(w_0, w_a)$ space. Others have argued that $(w_0, w_a)$ should be interpreted as `fitting parameters', meaning that one should take a model's predictions for the exact cosmological observables (in this case the raw distance measurements which are sensitive Eq.~\eqref{eq:hubble_fit}) and use the $(w_0, w_a)$ parameters to determine a best fit for those predicted observables. Regardless, the models considered here will still sweep huge regions of the parameter space, this footnote is just to highlight that there are a few different interpretations of this parameter space and the exact representation of it is non-unique.
See \cite{Wolf:2023uno, Wolf:2024eph, Shlivko:2024llw, Wolf:2025jlc} for further discussion.}

As discussed in \cite{Wolf:2023uno}, within the region of field space for which a quintessence field can serve as dark energy, the predictions between many distinct microphysical models are, both in principle and in practice, indistinguishable from each other in terms of their predictions for the equation of state and its resulting observables. For a brief concrete example, the typical hilltop model and the pseudo-Nambu--Goldstone Boson (pNGB) model can arbitrarily approach each other's predictions in ($w_0$, $w_a$) because, when their potentials are Taylor expanded, their leading order terms are identical. Further, it is these terms that describe the regime of field space responsible for the observed dark energy in the current epoch because dark energy given by an equation of state close to the cosmological constant value can only have undergone a fairly limited amount of evolution. Yet, for time-scales on the order of the life-span of the universe, their differences in microphysics would lead to either an abrupt recollapse of the universe in the case of the standard hilltop model because the potential eventually becomes negative \cite{Felder:2002jk}, or merely a peaceful end to further acceleration in the case of the PNGB model because this potential eventually stabilizes and oscillates around its minimum \cite{Frieman:1995pm}. Nothing less than our knowledge of the future fate of the universe is at stake here! 

Furthermore, in analogy with the single-field inflation paradigm, the theories of dark energy described above by the quintessence paradigm all fall within a common but simple framework: that is, they are all described by a single, minimally coupled scalar field with a canonical kinetic term and a potential function. Consequently, the ability for all of these distinct models to saturate the same observable parameter space is not artificially generated by engineering unrealistically complex or \emph{ad hoc} constructs. They are all on a relatively level playing field, described by the simplest imaginable way to build scalar field theories within general relativity on an FLRW cosmological background.

\section{Underdetermination}\label{Sec:underdetermination}

\subsection{Types of underdetermination}

The underdetermination of theory by evidence is undoubtedly a central pillar in the realism debates in the philosophy of science\footnote{While it is not necessary to subscribe to realism to be concerned about underdetermination (as will be discussed briefly later on this section), given that most physicists themselves are favorable to scientific realism in some form (e.g.\ see Jim Peebles' perspective on what he takes to be the physics community's operative philosophy of physics \parencite{Peebles:2024jdt}), we find realism to be an informative lens through which to analyze underdetermination in cosmology. It best captures the concerns of the physics community which largely sees physics as a project working towards a fundamental understanding of ``a unified mind-independent physical reality'' where our best theories are taken to be good approximations of this reality \parencite[p.~2-3]{Peebles:2024jdt}.}, where the familiar distinction between `transient'/`weak' underdetermination and `strong' underdetermination delineates the boundaries of our epistemic misgivings \parencite{Sklar1975-SKLMC, Quine1975-QUIOEE, Duhem1954-DUHTAA}. As the familiar story goes, there might be a number of theories competing to explain the available data; yet, they differ in their empirical predictions, which suggests that such underdetermination is transient and will be broken once further empirical data can be gathered. Far more epistemically worrying \emph{prima facie} is the possibility that there exist a number empirically equivalent theories that could never be distinguished from each other by any empirical data, but which also present distinct and conflicting ontological visions of the world. Here, we take empirical equivalence between theories $T$ and $T'$ to mean the \textit{exact} equivalence between the empirical substructures of every model $M$ of $T$ and $M'$ of $T'$ \cite{VanFraassenBas1980-VANTSI}. This strong underdetermination represents a serious challenge to those with realist predilections because it seems to undermine any firm basis for using science
to identify our ontological commitments.

However, the debate concerning the degree of epistemic threat posed by strong underdetermination has largely hinged on whether there are any truly compelling examples of such underdetermination. On the one hand, some philosophers have taken the threat seriously (e.g.\ \cite{Quine1975-QUIOEE, Jones1991-JONRAW, Kukla1993-KUKLLE, Earman1993-JOHURA, Wolf:2023xrv, Hoefer2020-HOESRW, 
acuna2021charting, mulder_read}) and pointed to, among other examples, alternative formulations of quantum mechanics, Newtonian mechanics, and general relativity to argue that there may be genuine instances of strong underdetermination. On the other hand, these examples have all generated a fair amount of skepticism, with skeptics dismissing such examples as artificial, and, for example, arguing that the theories in question are either notational variants of one and the same theory, or that the proposed `alternatives' are deficient in some obvious way (e.g.~\cite{Laudan1991-LAUEEA, Norton2003-NORMEU, Musgrave1992-MUSRAW, Stanford2001-STARTD-5}). \textcite[p.~20]{Norton2003-NORMEU}, in this context, has prominently argued that, in any case where we can tractably demonstrate empirical equivalence between two theories, ``we cannot preclude the possibility that the theories are merely variant formulations of the same theory'', and that this suggests that we should view purported instances of strong underdetermination with suspicion.

More recently, \textcite{Pitts2010-PITPUF} has identified a third form of underdetermination, dubbed `permanent underdetermination'. Rather than models sharing exactly equivalent empirical substructures as in the case of strong underdetermination, here the idea is that the models are technically empirically inequivalent, but nevertheless arbitrarily close in their empirical substructures. As an example, Pitts considers the approximate empirical equivalence of various massless theories in modern particle physics and gravitation research alongside their massive counterparts. That is, consider that $\left\{(\forall m) T_m\right\}$ is a family of related theories parameterized by mass $m$, whereas $T_0$ is the corresponding massless theory. $T_0$ and $\left\{(\forall m) T_m\right\}$ approximate each other arbitrarily closely in the limit $m \rightarrow 0$. So while $T_0$ may in principle be transiently underdetermined with certain members $T_i$ of the family, as long as $T_0$ remains viable it can \textit{never} be empirically distinguished from the larger family $\left\{(\forall m) T_m\right\}$. Crucially, ``the empirical equivalence is not merely approximate, and hence perhaps temporary; rather, the empirical equivalence is \textit{arbitrarily close and hence permanent}'' \cite[p.~271, our emphasis]{Pitts2010-PITPUF}.

This novel type of underdetermination is arguably far more interesting and compelling than strong underdetermination, if only for the reason that this type of underdetermination is immediately immune from the common charge that the theories in question are merely notational variants of each other. They plainly cannot be `one and the same' because they are empirically inequivalent and make different ontological claims; yet, there is also a precise sense in which they can never be distinguished from one another empirically. Furthermore, in addition to the aforementioned worries concerning our ability to appropriately judge ontological commitments, permanent underdetermination also generates epistemic worries further upstream as it undermines our confidence in science's ability to determine the best empirical descriptions of nature amongst obviously distinct alternatives. Consequently, and unlike strong underdetermination, cases of permanent underdetermination do not afflict \emph{only} scientific realists and are potentially of concern to anti-realists/instrumentalists as well. In other words, the worry is that it cuts us off from using empirical information to further refine the instruments we use to describe nature.\footnote{Our thanks to an anonymous reviewer for emphasising this point.}

\subsection{Permanent underdetermination in cosmology}

Up to this point, philosophical attention regarding underdetermination in cosmology has focused largely on allegedly strong underdetermination in large-scale spacetime geometry and topology \cite{Butterfield:2014twa, Belot2023-BELAEP-2, Ellis:2006fy, Manchak2009-MANCWK-2}, or stayed closer to transient underdetermination (implicitly and/or explicitly) and explored how various extra-empirical or methodological considerations might in the meantime influence matters of interpretation, theory-choice, or theory-pursuit given the (quite challenged) observational \emph{status quo} in the early universe or dark matter/energy \cite{WolfDuerr, Wolf:2022yvd, Dawid:2023mmv, AntoniouForthcoming-ANTTPO-14, Duerr:2023frq, Martens:2020lto, Azhar:2016mpg, Massimi2021-MASCBD, Smeenk:2017uof, Koberinski:2022zok, Allzen:2024wke, Jreige:2024ryw}.
However, this paper confronts the possibility that cosmology might well be plagued with \emph{permanent} underdetermination in the above sense, and indeed that this more pernicious underdetermination applies to distinct models within the \textit{same} theories/frameworks. The upshot is that cosmological modeling might already be hopelessly undetermined even before departing from the simplest ways of describing concrete cosmological observables in an expanding, perturbed FLRW spacetime.

To be a little more specific, the issue of permanent underdetermination in cosmology is the following. In the dark energy case, one can always find multiple distinct microphysical models which come arbitrarily close in their predictions of the parameters $(w_0, w_a)$.\footnote{To be clear, this applies regardless of whether or not the most recent indications from the data that dark energy might be dynamical hold up. If the data pulls back towards a cosmological constant, all the options are still on the table as all of the models discussed here (and many more) are all perfectly capable of mimicking a cosmological constant to produce $(w_0, w_a) \simeq (-1, 0)$. If the data continues to pull away from a cosmological constant, we may be able to eliminate $\Lambda$ as a viable candidate (an example of eliminative reasoning in this content \cite{Koberinski:2022zok}), but that would still leave a multitude of completely distinct dynamical possibilities on the table.} Likewise, in the inflation case, one can always find multiple distinct microphysical models which come arbitrarily close in their predictions of the parameters $(n_s, r)$. So, in both cases we have an apparent case of permanent underdetermination, and it is incumbent upon us to attempt to overcome this if we are to identify a specific cosmological model which is best apt to describe our universe.\footnote{One might claim that, having restricted \emph{ex cathedra} to single scalar field models of inflation and dark energy, there is no threat to scientific realism here, because all models agree on an ontology of a single scalar field $\varphi$. However, strict entity realism is not really what any realists (or physicists) we know are after. We do not only want to know whether some field exists, but we also want to understand its dynamical properties, how it interacts with other fields, etc. akin to how the standard model of particle physics has revealed the nature of bosons and fermions and how they interact to form the rich mosaic of microscopic world. Even if the only possibility was scalar field, the form of the microphysical models (as given by the potential $V(\varphi)$) can still differ significantly from model to model---so that aspect is still substantially underdetermined. Moreover, as we discuss below, we restricted to such single scalar field models principally to simplify our narrative---however, we expect the selfsame issues of underdetermination to persist when one considers other kinds of (non-single scalar field) models of which there are plenty, and in that case there is indeed the kind of ontological underdetermination about which our above interlocutor is worried.}\textsuperscript{,}\footnote{Permanent underdetermination in the cosmological case is somewhat different from Pitts' case because in the former there is no natural analogue of the massless theory. This is of course true---but we don't take the observation to detract from the points which we are seeking to make in this article.}

Before proceeding, it is worth pausing briefly to say just a few more words concerning the observational status quo and the diagnosis of permanent underdetermination. Typically, when analyzing potential instances of strong or permanent underdetermination, the implication is that the underdetermination holds with respect to all possible observations. Here we have identified and focused on the primary observables relevant to testing and constraining dark energy and inflation models. Is it possible that there are other empirical factors that could come into play that might lead to the conclusion that these are not examples of permanent underdetermination?

In our view, the answer is almost certainly `no'. The first thing to be said is that our empirical access within cosmology as a whole, and to the early and late-time universe physics that we attempt to model with inflation and dark energy in particular, is incredibly limited. With inflation, the actual physics occurs at an epoch and at energy scales to which we have no direct empirical access. We are limited to gathering relic statistical imprints produced by the actual physical process---well after the fact and only once the universe has cooled enough to allow photons to stream freely. While we have some small measure of direct empirical access to dark energy because we are living through this epoch at present, this empirical access is limited to just a few basic kinds of measurements that chart out the expansion history or growth of cosmic structure on the largest scales in the universe. As discussed in detail by \textcite{FWR}, these data points are useful (but blunt) instruments that give us some insight into the bulk properties of these energy components' fluid-like descriptions, but leave details of their microstructure massively unconstrained. This is similar to how measuring the viscosity of a fluid might give us some insight into its properties, but utilizing only this information, there is very little we could say about its detailed molecular or atomic structure. Given this state of affairs, 
it is almost certain that observables like those identified here will forever remain the only relevant observables that one can use to make any substantive statements about the physics of inflation or dark energy, and these observables only give (at best) a limited glimpse at what the underlying microphysical structure might be. 

The second thing to say is that, while there are some other observational parameters that can be constrained beyond $(w_0, w_a)$ and $(n_s, r)$ that, under some very particular circumstances, might come into play to tell us something about dark energy or inflation that the primary observables are not themselves able to, 
there are very good reasons to believe that such observables will not affect this diagnosis of permanent underdetermination. 

Two reasons for this are as follows. First, as discussed by \textcite{FWR}, most other potential observables discussed in these contexts as possibilities would necessarily be far fainter and more poorly constrained when compared with the primary observables as they have not yet been detected. Second, both the single-field inflation and quintessence paradigms represent essentially the simplest way of building scalar field theories relevant to cosmology, and they both happen to offer empirically adequate and viable descriptions of the regimes which they purport to describe. These other possible observables represent telltale signs of highly exotic physics that go beyond these simple frameworks. For example, cosmologists also consider the possibility of finding non-Gaussian signatures in the primordial density perturbations. However, it is known that simple inflation models such as the ones discussed here produce unobservably small non-Gaussianities \cite{Martin:2013tda}. Observations of primordial non-Gaussianity would necessitate a move to more complicated models, such as those with non-canonical kinetic terms or with sharp features in their potential functions \cite{Chen:2010xka}. Similarly, cosmologists have been looking for evidence of fifth forces that could conceivably show up in solar system tests or in the growth of cosmic structure. If evidence revealing such effects was confirmed, it would necessitate moving away from the simple quintessence framework and towards true modified gravity theories such as scalar-tensor theories with a non-minimal coupling to the Ricci scalar \cite{Joyce:2016vqv}.\footnote{See e.g.~\cite{Wolf:2024stt, Ye:2024zpk} for some recent discussion of how non-minimally couple scalar-tensor theories might alleviate some perplexing aspects of current cosmological data.} 
In either case, further observational signatures beyond the main observables described here point us towards substantially more exotic physics that \textit{requires} the introduction of more parameters and more complicated interactions. Given that we have permanent underdetermination at the simplest level of empirically adequate description, we have every reason to expect that the underdetermination problem would be even worse if observations required that we adopt more complicated frameworks with larger parameter spaces. 

To sum up: barring some as-yet unconceived revolution that would fundamentally change the kind of empirical access we have to cosmological phenomena, it is very likely that both inflation and dark energy are permanently underdetermined \cite{FWR}. 
Due both to the inherent empirical limitations and access within cosmology, it is almost certainly the case that these will remain the primary observations for making any substantive empirical statements about inflation or dark energy. While some other possible observational signatures beyond these are conceivable if inflation and/or dark energy are significantly more exotic than conceived here, detecting such signatures would likely make the problems of permanent underdetermination even worse for the reasons mentioned above. 

Ultimately, we want to get as close as we can to the underlying physical theory that describes the evolution of the universe. While this is of course a tremendously ambitious goal, finding ways to break or lessen the underdetermination certainly has the potential to make a positive contribution in this direction. Currently, physics is inundated with  hundreds (if not thousands) of `toy' models and variegated theoretical proposals for inflation and dark energy. A strong justification for pursuing strategies to break or weaken this underdetermination is to single out privileged descriptions of the relevant physics, and thereby identify redundancies, enhance understanding, and sharpen the heuristics used for investigating cosmological phenomena in the hopes of moving closer to this goal.

\subsection{Responses to underdetermination}

What responses are available when presented with cases of permanent underdetermination?
To explore an answer to this question, we can avail ourselves of a (suitably modified) taxonomy of possible responses to strong underdetermination given by \textcite{LeBihan2018-LEBDAO}. Of these, three strategies stand out as potentially having relevance for permanent underdetermination:
\begin{description}
    \item[Discrimination:] Preferentially discriminate in favor the ontological claims of one theory amongst the underdetermined alternatives.\footnote{E.g.\ consider that one might break the underdetermination between various different formulations of electromagnetism in favour of the fibre bundle formulation both of grounds of (a) ontological parsimony and (b) expressive power (since this formulation still admits a variational principle etc.).}
    \item[Common Core:] Break the underdetermination by moving to a new interpretive framework. The new framework is obtained by isolating the `common core' that is shared among the underdetermined alternatives and then interpreting this shared common core as a distinct, ontologically viable theory of its own.\footnote{E.g.~see \cite{DeHaro2021-DEHOSA, March:2023trv} for applications of the common core approach in response to Newtonian-themed instances of strong underdetermination, where Maxwell gravitation/spacetime could be argued to be the common core.}
    \item[Overarching:] Break the underdetermination by developing a new (potentially richer) theoretical structure which subsumes the original underdetermined theories.\footnote{E.g.~see \cite{Muller1997-MULTEM, Muller1997-MULTEM-2} for discussion on how matrix and wave mechanics were synthesized into the now-standard formulation of quantum mechanics based upon Hilbert spaces.}
\end{description}

While these strategies have all frequently been pursued in the context of strong underdetermination, they might also be applied profitably in response to cases of permanent underdetermination. Of the three, the discrimination approach is fit for purpose as is and requires no modification. There evidently can be reasons to prefer one theory over another in cases of permanent underdetermination, including (but not limited to) super-empirical virtues (e.g.~simplicity, coherence, predictive novelty, etc.), explanatory power, and the lack (or presence) of theoretical structures deemed pathological.

On the other hand, applying the common core and overarching approaches to permanently (as opposed to strongly) underdetermined theories requires a little more thought. Begin with the common core strategy: here, one is guided by the need to construct some  weaker (i.e.\ structurally more impoverished) theory which is nevertheless empirically equivalent to the original underdetermined theories. As such, it is not so obvious how to identify the common core when empirical equivalence fails, as is indeed the case in instances of permanent underdetermination. One strategy here would be to focus only on empirical equivalence \emph{in some domain}, and proceed from there.

When it comes to the strategy of building an overarching theory, the situation is this.
Overarching theories, such as M-theory subsuming various superstring theories or quantum mechanics subsuming matrix and wave mechanics, exhibit a richer solution space than the theories they encompass, which is not terribly surprising considering that such a framework by necessity must be more general in some sense. Of course in the case of permanent underdetermination, the new ontological framework stemming from the common core or overarching strategies will necessarily not be precisely equivalent to the underdetermined theories as they are not precisely equivalent to each other. Yet, that notwithstanding, nothing would seem to preclude one from following the `overarching' strategy when faced with permanent underdetermination.

Now, when considering both the common core and overarching approaches, a point made by \textcite{LeBihan2018-LEBDAO} in the context of strong underdetermination bears stressing: simply constructing a new theory (whether a common core theory or an overarching theory) does not \emph{per se} ameliorate philosophical problems of underdetermination---in fact, there is a clear sense in which developing some new theory makes the situation worse! As such, these strategies must be supplemented with further philosophical reasoning (e.g.\ reasoning in terms of parsimony or explanation or unification) in order to justify treating the newly-developed theory as preferred, and thereby to overcome the case of underdetermination under consideration.\footnote{Note that the common core approach places weight upon parsimony, whereas `overarching' strategies seem in general to place more weight upon  unification.} When, later in this article, we speak of `breaking the underdetermination' between theories, we mean that, conditional on the adoption of reasoning such as the above, the symmetry between the elements of that class of underdetermined theories is broken, in the sense that this reasoning is deployed in order to afford greater significance to the ontological claims of one theory in the class versus those of the others. This point continues to stand when these strategies are brought to bear on cases of permanent underdetermination, which is our concern here.

\section{Effective field theories}\label{sec:eft}

\subsection{The EFT paradigm in physics}

Effective field theories (EFTs) are ubiquitous in modern physics. The essence of the EFT paradigm is this: we take some target system which is in some sense and to some degree isolated from external influences, and we are interested in providing a description of this target system up to a level of precision which makes sense relative to the physics of the system as compared with that of the environment and of the relevant measuring devices (this could involve a comparison of energy scales, or of length scales, or of something else, depending upon context). So, there is a scale-relativity built into the EFT paradigm. Often, this scale-relativity is indeed built into the model explicitly: one defines a power counting parameter $\delta$ such that quantities can be calculated to some order in $\delta$; relative to a given modelling context, terms sufficiently high order in $\delta$ will be negligible.
%\footnote{For a recent articulation and defence of this way of thinking, see \parencite{LinnemannReadTeh}.}

It is by now well-recognised that both the Standard Model of particle physics and general relativity can be understood as EFTs. As \textcite[p.\ 241]{Burgess} writes on the latter:
\begin{quote}
    From this point of view the Einstein--Hilbert action should not be regarded as being carved by Ancient Heroes into tablets of stone; one should instead seek the most general action built from the spacetime metric, $g_{\mu\nu}$, that is invariant under the symmetries of the problem [...] organised in a derivative expansion.
\end{quote}
As we'll explore later, the actions which have been offered in inflation and dark energy models are also highly plausibly understood as being those associated with EFTs---and, indeed, this offers some novel possibilities for tackling underdetermination in cosmology in general. Before we get to that, though, a little more on the connections between the EFT paradigm on the one hand and underdetermination of theory by evidence in the other.

\subsection{EFTs and underdetermination} Suppose now, following \textcite{Polchinski2017-POLDOF}, that one has some high-energy theory which admits of multiple distinct perturbative expansions---expansions, indeed, which might agree up to some order (in the power counting parameter $\delta$) but diverge thereafter.\footnote{For \textcite{Polchinski2017-POLDOF}, such a situation is definitional of a `duality' in physics.} Then, associated with the high-energy theory will be multiple distinct low-energy theories---theories which, indeed, might be approximately (but not exactly) empirically adequate in some domain. In a specific situation in which there is a large number---perhaps even an infinity---of such theories (a situation illustrated by \textcite[\S2.5]{Polchinski2017-POLDOF} in the context of the Montonen--Olive duality), this plurality might even give rise to a case of permanent underdetermination!

So, the EFT paradigm can (at least in some cases) afford a means of understanding the \emph{origins} of cases of permanent underdetermination such as those encountered in modern cosmology. But as we'll discuss in the next section, it also affords a novel way of thinking about various ways in which such cases of underdetermination might be resolved.

One last word on this: a precondition for deploying the EFT paradigm in order to overcome apparent cases of permanent underdetermination in cosmology is that one can be a scientific realist about EFTs at all: given that (by definition!) EFTs are effective only in some domain, and might break down thereafter, one might worry about such an approach. For an engagement with authors who voice such concerns, and for a compelling corrective that EFTs can and should be interpreted realistically, we refer the reader to the work of \textcite{Williams2019-WILSRM}, which we endorse wholeheartedly going forward, and which is quite naturally understood as being part of a broader recent movement in the philosophy of science towards regarding ontology as being `scale-relative' \cite{Ladyman2007-LADETM}, and towards thinking in particular that one's ontological commitments in a given physical context should be given by the mathematics which best describes the physical goings-on in that context (see, in particular, the `mathematics-first structural realism' of \textcite{Wallace2022-WALSSR-3}). Our discussions in this article are properly situated within this school of thought. Of course, this isn't to deny that the `effective realism' programme has its critics---see e.g.\ \textcite{Ruetsche2018-RUERGR, Reutsche2024, McKenzie_forthcoming_Prospects, McKenzie2024_NoGrounds}---but here is not the place to discuss those more general concerns.

\section{Addressing permanent underdetermination in cosmology}\label{Sec:responses_in_cosmo}

\subsection{Responses to permanent underdetermination in inflationary models} The situation \emph{vis-à-vis} permanent underdetermination and inflation is as follows. It seems to be the case that given a pair ($r$, $n_s$), which represents the primary cosmological observables relevant to an inflationary epoch in the early universe, there will always be a plethora of distinct microphysical models that can generate predictions for ($r$, $n_s$) that are arbitrarily close to each other. Thus, we have an instance of permanent underdetermination. As we will be interested in exploring the extent to which we can successfully break this underdetermination, whether by identifying a privileged ontology of one of the theories or by finding some new ontology in which to embed the underdetermined theories, it is worth briefly reflecting on the ontological posits of the standard inflationary paradigm.

Standard inflation can be described succinctly as being given by models of the form $\langle M, g_{\text{FLRW}}, \Phi_i, \varphi \rangle$, where $M$ is a four-dimensional differentiable manifold, $g_{\text{FLRW}}$ is the FLRW metric on $M$, $\Phi_i$ represent other matter fields (e.g.\ standard models fields, dark matter, etc), and $\varphi$ represents the inflaton field. As there are many distinct microphysical models, these will all pick out distinct dynamical possibilities from amongst this set. These dynamically possible models will then be given by $\langle M, g_{\text{FLRW}}, \Phi_i, \varphi_V \rangle$, where $\varphi_V$ denotes a specific microphysical model of inflation determined by the particular potential function $V(\varphi)$ that describes it. Furthermore, these dynamical possibilities all obey dynamics given by the Klein-Gordon equation in an FLRW background,
\begin{equation}\label{eq:ScalarEOM}
    \ddot{\varphi} + 3 H \dot{\varphi} + V'(\varphi) = 0,
\end{equation}
with $V'(\varphi) = dV/d\varphi$. The solutions for $\varphi$ will of course depend upon the particular inflation model as the functional form of $V$ will dictate the model-specific dynamics of the scalar field. These dynamics then get fed into $H$, which determines the dynamical trajectory of the universe itself through its impact on the scale factor $a$. With this in mind, we identify two plausible strategies that can be deployed in response to permanent underdetermination in inflation: the discrimination approach and the overarching approach. 

Discriminating would involve favoring the ontological claims of some particular model out of all those considered. Given our background knowledge from the Standard Model of particle physics, it turns out that there is a uniquely privileged candidate: Higgs inflation, denoted by $\langle M, g_{\text{FLRW}}, \Phi_i, \varphi_H \rangle$. As the only fundamental scalar field that has been empirically verified, at first glance the Higgs seems to have the properties we are after: it is a scalar field that permeates all of space in order to contribute to the universe's energy density and it has a flat region in its potential. If it were concluded that the standard model Higgs, before it reached the minimum of its potential that it now occupies, produced an inflationary epoch consistent with observations, there would be an open-and-shut case for discriminating in favor of $\varphi_H$. The resulting consilience, coherence, and parsimony with respect to the most precise, empirically verified, and fundamental theory that physics is in possession of would be so overwhelming that it is hard to imagine there would be any desire for physicists to investigate the other many hundreds (literally) of `toy' models that have been considered. However, this tantalizing scenario ultimately does not work; there are excellent constraints on the parameters of the standard model Higgs, and the observed value of the self-coupling constant and the Higgs mass produce amplitudes for density perturbations many orders of magnitude larger than those which are actually observed \cite{Bezrukov:2007ep}.

While the Higgs field, understood exactly according to the Standard Model of particle physics, is not a viable inflation candidate, there does perhaps remain a way in which to salvage a discrimination-type argument in its favor. As discussed in \cite{Bezrukov:2007ep, Martin:2013tda}, at very high energies, renormalizing a scalar field generally creates a non-minimal coupling between the scalar field and the Ricci scalar of gravity because quantum corrections typically introduce such terms in the effective action. With these considerations in mind, it has been shown that Higgs inflation with a non-minimal coupling can produce inflation in excellent agreement with observations with a nearly scale-invariant spectrum and $r\sim10^{-2}$. If future observations were to indicate strong agreement with these predictions, then there would be a very strong argument for discriminating in favor of Higgs inflation as similar reasoning to that detailed above would still apply. Higgs inflation with a non-minimal coupling would be strikingly cohesive with the Standard Model of particle physics, and the only new physics required by such a scenario would be that which is already expected as a natural consequence of renormalizing scalar fields in a curved spacetime background.\footnote{While this argument can be made in compelling fashion at the level of theory virtues (e.g.\ simplicity, coherence, predictive novelty, etc.)\ \cite{Kuhn1981-KUHOVJ, Schindler2018-SCHTVI-5}, one can also imagine making such an argument from the perspective of the meta-empirical arguments given in \cite{Dawid2013-DAWSTA}.} At that point, it would be difficult to argue that other inflationary models should be taken as serious competitors.\footnote{Higgs inflation with a non-minimal coupling is also closely related to Starobinsky inflation, another model with strong theoretical motivations. See \cite{Martin:2013tda} for detailed discussion.} This scenario would also be ideal for pursuing further questions in cosmology or high energy particle physics given that many of the various couplings and interactions with other particles are already known quantities. 

Of course, there is no guarantee that this scenario will play out. Observations might instead favour another region of parameter space, or the upper bounds on $r$ might get pushed below observational sensitivities. Another clear approach that can be distilled from the literature is strongly analogous to the overarching approach and is explicitly due to some physicists' stated desires to work in an `agnostic' or `model independent' way given the lack privileged microphysical model. The strategy is then to embed the inflation paradigm in an EFT.\footnote{There are numerous conceptual issues with applying the EFT framework to inflation and cosmology more generally. Briefly, the usual separation of scales that is present in other EFT applications does not seem to hold in the same way in cosmology. Here, we set aside these issues and take for granted that these methods can be applied. See \cite{KoberinskiForthcoming-KOBAT, KoberinskiForthcoming-KOBEAT-2} for deflationary views from the philosophy literature and \cite{Martin:2000xs} for a physics formulation of the so-called Trans-Planckian problem which looms large in these discussions. See \cite{Burgess:2020nec} for a rebuttal from the physics literature and \cite{Wolf:2022yvd} for a philosophical analysis of the heuristic value of biting the bullet and accepting this breakdown of scales.} There are a few approaches \cite{Cheung:2007st, Weinberg:2008hq, Azhar:2018nol, Burgess:2017ytm}, but that of \textcite{Cheung:2007st} is arguably the most well-known. 

Here, the authors apply the EFT-building philosophy to the problem of inflation. That is, given that the main observable constraints are directly sensitive to scalar fluctuations, they construct the effective action at the perturbative level for these inflationary scalar fluctuations with ``the lowest dimension operators compatible with the underlying symmetries'' \cite[p.~1]{Cheung:2007st}. That is, the physical situation in which we are interested is the description of scalar fluctuations around a quasi-de Sitter background. Here, the relevant symmetries are spatial diffeomorphisms and time diffeomorphisms, but the scalar field acts as a `clock' that breaks the time-translation symmetry which the de Sitter background would otherwise have had (hence, `quasi'-de Sitter). Schematically, such a theory can be written in the following way \cite[Eq.~3.32]{Burgess:2017ytm}:
\begin{equation}
\mathcal{L} = \frac{M_p^2}{2} R-\alpha(t)-\beta(t) g^{00}+\frac{1}{2} M_2^4(t)\left(g^{00}+1\right)^2+\frac{1}{3!} M_3^4(t)\left(g^{00}+1\right)^3+\cdots.
\end{equation}
Here, the theory has been written in the so-called `unitary gauge' where the scalar degree of freedom is absorbed into the metric $g$. The first term represents gravity through the Einstein--Hilbert term, while the next two terms encode the unperturbed dynamics of the background spacetime and scalar field. The higher order terms can be built out of the temporal part of the metric $g^{00}$, the extrinsic curvature $K_{\mu \nu}$, the Riemann tensor $R_{\mu \nu \rho \sigma}$, etc.\ (see \cite[Appendix A]{Cheung:2007st} for details). While in principle the coefficients in front of the various terms represent arbitrary functions of time, specific choices for these functions will correspond to familiar inflationary models. For example, the phenomenology of the simplest inflation models discussed here can all be understood to be contained within the first three terms here, whereas the higher order terms describe the phenomenology that results from deviations from this paradigm (e.g.\ the action for standard slow-roll inflation is given by the choice $\alpha=V(\varphi)$, $\beta=\frac{1}{2}\dot{\varphi}^2$, and all other functions parameterizaing the higher order terms are set to zero). The higher order terms might capture higher order effects such as non-Gaussianities which we would expect to derive from e.g.\ non-standard or higher order kinetic terms.

What we have here represents a clear-cut case of applying the overarching approach. As an analogy, consider well-known examples that have been identified in the literature as exemplifying this strategy, which include (to repeat from above) embedding the various superstring theories within the framework of M-theory, or embedding matrix and wave mechanics into what is now consider to be `orthodox' quantum mechanics \cite{LeBihan2018-LEBDAO, Muller1997-MULTEM, Muller1997-MULTEM-2}. The distinctive feature of this strategy is that the underdetermined theories have been unified such that they can be understood as different facets of the overarching theory that subsumes them. This is exactly what has been done here. That is, the above inflationary EFT represents the most general framework compatible with the most basic physical assumptions of inflation (quasi-de Sitter expansion in a perturbed FLRW background), and the various microphysical inflationary proposals correspond to particular choices for $\alpha$, $\beta$, and the functions parameterizing the higher order terms.\footnote{That said, there are still important differences between these cases---most obviously, von Neumann's work in quantum mechanics obviously amounted to more than merely constructing a theory with more free parameters, which is essentially what we have in the inflation case.} However, it is also important to emphasize that this framework is far more general than the simplest versions of the inflation paradigm, and can accommodate much more exotic physics as particular realizations of the various EFT parameters. 

What is the fundamental ontology posited by this framework? The ontology still consists of a scalar field, but the scalar field is now frequently denoted $\pi$ to distinguish it from the standard inflaton field $\varphi$. The inflationary EFT can be described by models of the form $\langle M, g_{\text{FLRW}}, \Phi_i, \pi \rangle$ and the dynamics for $\pi$ come from the very long and cumbersome EFT action schematically introduced above. While we are still working with a scalar field, there are some changes in its interpretation. $\pi$ is now interpreted as a Goldstone boson that results from the spontaneous breaking of time-translation symmetry, which generates some level of analogy with other dynamical systems in particle or condensed matter physics that exhibit spontaneous symmetry breaking. 

However, as noted in \S\ref{Sec:underdetermination}, the existence of an overarching theory does not by itself break the underdetermination. There is a further interpretive move that has to be made to justify the overarching framework over its various constituents. While what such a justification looks like will obviously be context dependent, as discussed earlier, what we are really looking for is an argument that would uniquely privilege one of these theories, with the ultimate goal being to develop the best theoretical description that can predictively account for cosmological phenomena and provide good explanations for (or even resolve) the scientific questions that we are interested in.

Unfortunately, in contrast with Higgs inflation, such a justification for the overarching theory is lacking. The EFT of inflation is \textit{only} valid for the period of inflation itself \cite[p.~17]{Cheung:2007st}. If there was an inflationary period in the early universe, we know that inflation had to end at some point and that a subsequent period of reheating is needed to describe how the inflaton decayed and the universe was populated with the mass-energy content observed today (i.e.\ the matter fields $\Phi_i$). The specific microphysics that dictates the nature of these particle interactions is relevant to these processes. In other words, the $\varphi_V$ component of $\langle M, g_{\text{FLRW}}, \Phi_i, \varphi_V \rangle$ is relevant for understanding the $\Phi_i$ component once inflation has ended. And working with $\pi$ obscures these links. Consequently, the totality of the physics relevant to the problem ensures that this overarching theory does not remove the need to explore and refine specific microphysical models. Furthermore, this particular EFT approach offers only limited epistemic value for understanding the microphysics of inflation. This is because it essentially offers a very general parameterization of possible physical effects that can result from a scalar degree of freedom.
This is not to deny that there is significant pragmatic value in the overarching theory in that it ``allows a relatively model-independent survey of what kind of observables are possible at low
energies, without having to go through all possible microscopic models beforehand'' \parencite[p.~86]{Burgess:2017ytm}. This can give us some insight into the general classes of inflationary models that might fit well with the data, but will not by itself offer any kind of perspicuous interpretation in terms of a particular microphysical model of inflation. That is, actual candidate microphysical theories will occupy different parts of the parameter space that this EFT defines as such microphysical theories generally do not instantiate each and every term that is allowed by the symmetry principles. In other words, the EFT itself does not present itself as a viable microphysical structure which could be the source of inflation, which is ultimately what we are after. While this EFT approach is no doubt valuable for describing the inflationary epoch and provides a very useful and informative tool that can help to constrain future model building efforts, it does not meaningfully defuse underdetermination concerns.

\subsection{Responses to permanent underdetermination in dark energy models}
The situation for dark energy can be set up in much the same way as for inflation. We have a plethora of microphysical models of the form $\langle M, g_{\text{FLRW}}, \Phi_i, \varphi \rangle$. The dynamical possibilities, which are a subset of these models, then correspond to $\langle M, g_{\text{FLRW}}, \Phi_i, \varphi_V \rangle$, where $\varphi_V$ denotes a specific microphysical model of quintessence that obeys the dynamics that follow from its potential function $V$ and the solutions to Eq.~(\ref{eq:ScalarEOM}). In this case though, the scalar field $\varphi$ is not totally dominant but rather competes with the already-existing matter fields $\Phi_i$ for influence over the dynamics of the universe, which generally makes these dynamics more complicated. Yet, there are many distinct microphysical models which give arbitrarily close predictions for ($w_0$, $w_a$) and are thus indistinguishable from each other. Similarly, one response to this situation mirrors the inflationary case. There is an almost identical EFT approach to dark energy that has been developed and applied over the years \cite{Gubitosi:2012hu} (i.e.,\ write down all the terms in the action that the symmetries of the problem allow and constrain the free functions that parameterize those terms); however, this is not the only option as one can motivate a different kind of effective field approach. Below we will argue that there is a straightforward application of the common core strategy available in response to permanent underdetermination in dark energy.

Depending on the problem of interest, it is often the case that physicists consider a Taylor expansion of the potential $V$ to some order in $\varphi$ when working with scalar field cosmological models \cite{Dutta:2008qn, Boyle:2005ug, Wolf:2023uno, Wolf:2024lbf, Kallosh:2002gg, Chiba:2009sj}. In other words, any arbitrary, analytic potential can be represented by a series expansion:
\begin{equation}\label{eq:scalar_field_expand}
V(\varphi)=V_0+\left.\frac{d V}{d \varphi}\right|_{\varphi=0} \varphi+\left.\frac{1}{2} \frac{d^2 V}{d \varphi^2}\right|_{\varphi=0} \varphi^2+\left.\frac{1}{6} \frac{d^3 V}{d \varphi^3}\right|_{\varphi=0} \varphi^3+\cdots.
\end{equation}
While this is not exactly the same as the EFT philosophy pursued in the inflation case where the authors used symmetries to write down the most general theory under the given physical assumptions, it is still an EFT in the sense that it is focusing on the scale-relative effects of a general scalar field potential.
This is particularly interesting in the context of the dark energy problem due to the material facts with which we are confronted. The universe has only recently entered a period of accelerated expansion that has been found to be either indistinguishable from, or incredibly close to, a cosmological constant depending on the data considered. 
All of the empirical facts on the ground are telling us that $w_{\mathrm{DE}} \simeq -1$ over the period of comic history to which we have robust empirical access. If dark energy is indeed driven by some scalar field within this general framework, this indicates that the field excursion will be small and that the dominant contribution will come from the constant part of Eq.~(\ref{eq:scalar_field_expand}), whereas the (small) deviations from the value predicted by a cosmological constant will necessarily be encoded in and dominated by the next-to-leading order term in the expansion. 

What does this term look like? For a large number of scalar field potentials, such as those whose functional forms are even or which have a critical point about the point at which the expansion is taken, the linear term in Eq.~(\ref{eq:scalar_field_expand}) automatically vanishes because the first derivative $V'$ is zero, leaving the quadratic term as the next-to-leading order contribution. This includes several well-known potentials such as hilltop potentials, the quadratic potential, axions, pseudo-Nambu--Goldstone bosons, Gaussians, various supergravity-motivated potentials, etc.,\ all of which look identical in this regime and can be described accurately by an energy scale $V_0$ and a quadratic term $V'' = m^2$ \cite{Kallosh:2002gg, Dutta:2008qn, Wolf:2023uno}, where we have now identified the second derivative of the scalar field potential as a mass term (more on this soon). What about potentials for which the linear term does not automatically vanish such as the frequently deployed exponential potential (which is often motivated by string theory considerations)? It turns out that even here, one can perform a field redefinition for the scalar field in order to eliminate the linear term and provide an equivalent description given by the rescaled field with a next-to-leading order quadratic term \cite{Wolf:2023uno}. The upshot is that, in the regime of field space where scalar field physics can describe dark energy, a tremendous number of the most widely used and theoretically well-motivated potentials can all be characterized to an excellent approximation with the same functional form given by
\begin{equation}\label{eq:effectivev0m2}
    V(\varphi) = V_0 \pm \frac{1}{2}m^2\varphi^2.
\end{equation}
Furthermore, this functional form happens to have the dynamical freedom mentioned earlier that allows it to saturate huge swathes of the observable $(w_0, w_a)$ parameter space. One the one hand, when $V''>0$ the dark energy equation of state has been found to evolve according to highly universal behavior characterized by slow, linear evolution \cite{Scherrer:2007pu, Wolf:2023uno}. While, on the other hand, when $V''<0$ the dark energy equation of state can evolve incredibly rapidly in a sharp, highly non-linearly manner depending on the choice of model parameters and initial conditions; this allows it to sweep over the observable parameter space \cite{Dutta:2008qn, Wolf:2023uno, Shlivko:2024llw}. This is due to the resulting effects on the parameter $w_a$, which captures the time variation of the equation of state. And finally, when $V'' (m^2) \rightarrow 0$, the model recovers the cosmological constant. 

In other words, this single functional form can account for the phenomenology associated with all dark energy models that fall under the umbrella of a single, canonical, minimally-coupled scalar field. The relevant scales and phenomena themselves seem to single out this kind of effective description for the physics. Furthermore, the fact that all of these distinct models can be understood to agree on this effective description of the physics makes this analogous to the common core strategy described in \S\ref{Sec:underdetermination}. That is, for every distinct microphysical dark energy model of the form $\langle M, g_{\text{FLRW}}, \Phi_i, \varphi_V \rangle$, there is an equivalent description (to arbitrarily close empirical precision) given by a model of the form $\langle M, g_{\text{FLRW}}, \Phi_i, \varphi_{(V_0, m^2)} \rangle$. The common core approach would then implore us to adopt this description, given in terms of an effective mass and energy scale, as it has been isolated by determining which aspects of the ontology are mutually agreed upon by all of the underdetermined models.\footnote{Cf.\ the arguments adduced in favour of Newton's law of universal gravitation in the face of (permanently) underdetermined alternatives in the hypothetical scenario contemplated by \textcite[\S1]{Ruetsche2018-RUERGR}. (Note that Reutsche does not use the terminology `permanent underdetermination', but her case study indeed seems to fit that mould.)}

As before, however, the mere existence of a viable common core does not by itself break the underdetermination. Further argumentation or interpretation is needed in order to justify the common core theory as successfully breaking the underdetermination. One clear justification takes the form of a `robustness argument' in favour of the common core of the underdetermined models: since the common core features in the plurality of underdetermined models (and is robust in that sense), we have some heightened degree of confidence that this common core offers explanatory value within the problem context of dark energy physics beyond the individual merits of each specific quintessence proposal viewed in isolation. To give an example, if one or many of the proposals within this family were shown to have theoretical pathologies that rendered them unsuitable candidates, the common core theory would still remain a viable construct. In other words, the common core is a robust feature of this family of dark energy proposals in the sense that it offers a unique description which all of these models flow to within the regime of interest. This robustness establishes the potential explanatory viability of the common core in a highly reliable manner, which by itself confers additional pursuit-worthiness to it on explanatory grounds. 
For discussion of such arguments in the context of a search for a quantum theory of gravity, see \textcite{Linnemann2020-LINNRA-2}.\footnote{Cf.\ \emph{divide et imperia} strategies discussed by \textcite{Ruetsche2018-RUERGR}.}

Another flavour of justification that often shows up in the context of adopting a common core theory over its rival description involves appeals to parsimony: if there is excess, idle structure in our ontology, then it is well-advised not to take such structure seriously when articulating one's roster of ontological commitments.
In the case of the permanent underdetermination of dark energy models, a justification exactly identical to the above isn't available because all of these theories share roughly the same basic ontological structure; i.e.\ there is some spacetime metric, matter fields, and a dark energy scalar and it's not obvious that there is any dramatic Occamist gain which results from moving to $\langle M, g_{\text{FLRW}}, \Phi_i, \varphi_{(V_0, m^2)} \rangle$ if parsimony is construed as ontological parsimony (the sheer quantity of entities or kinds of a particular entity). Yet, parsimony need not be exclusively construed in this way. In addition to ontological parsimony, there is also syntactic parsimony, which refers to the parsimony of the theory's structure, particularly in terms of the number and complexity of its assumptions, variables, or formal/mathematical elements \cite{Schindler2018-SCHTVI-5}. 

Here, the effective description really shines. The familiar mass/quadratic term leads to linear equations of motion which are formally equivalent to those of a damped harmonic oscillator when $m^2 > 0$, or a system exhibiting an exponential instability within this regime when $m^2<0$ (which also has many classical analogues). This means that, contra most scalar field potentials considered in the literature, the theory given by $\langle M, g_{\text{FLRW}}, \Phi_i, \varphi_{(V_0, m^2)} \rangle$ leads to Eq.~(\ref{eq:ScalarEOM}) having either shockingly simple analytic solutions or very manageable numerical solutions depending on the exact context. Of course, this generates insight into parameter dependencies, increases computational speed and tractability, and facilitates further predictive power (see e.g.\ \cite{Wolf:2023uno, Dutta:2008qn} for specific examples where this has been leveraged in this problem-context). There is also arguably a significant gain in understanding to be had as this theory allows us to import our pre-existing insights (both quantitative and qualitative) into a new application. We are just dealing with a field that possesses the property of mass, which is arguably the kind of physics that we have most epistemic control over at both the classical and quantum level as mass is simply a known intrinsic property of fields that quantifies their resistance to motion. This theory then lends itself to a familiar, perspicuous interpretation of the ontology that isn't always available if one is working with some highly exotic field that may have been introduced with dubious or speculative physical motivations in mind. Despite all of these dark energy theories being similar in terms of ontological parsimony, $\langle M, g_{\text{FLRW}}, \Phi_i, \varphi_{(V_0, m^2)} \rangle$ is clearly privileged in terms of its syntactic parsimony, for both pragmatic and epistemic reasons. 

Another factor which speaks in favour of the common core theory in this case has to do with its unification of all the various alternative microphysical models. Rather than painstakingly investigating each model individually, one can now investigate the whole family of models under their effective description in one go. This has been exploited in \cite{Wolf:2024eph}, where the authors were able to obtain constraints on the entire family of models through utilizing the effective description in terms of $V_0$ and $m^2$. Among other things, this allows one to directly glean information concerning the likelihood of the common core model parameters (that again captures the whole family of theories) when confronted directly with cosmological data. There it was shown that in light of the recent DESI data which favors a time evolving dark energy equation of state, models with $m^2 < 0$ are favored in terms of their likelihood over models with $m^2 \simeq 0$ or $m^2 > 0$, which provides some small measure of evidence for the detection of a `negative' cosmological scalar field mass (there are several important nuances to this statement that we are eliding over---see \cite{Wolf:2024eph} for more details). 

This reflects a model-agnostic approach to this general class of dark energy theories that allows one to evade the difficult and time-consuming task of investigating each and every distinct potential that can be dreamt up. Yet, if one, for some reason (maybe due to some more fundamental interest in a particular model(s)), did not want to be model-agnostic, this is useful here too. Such a unified description facilitates a like-to-like comparison of different theories which are known to occupy certain regions of the ($V_0, m^2$) parameter space using a common language in terms of the same parameters (e.g.\ the typical exponential model which has $m^2>0$ as opposed to, say, an axion model with $m^2<0$). Furthermore, one can always map between the parameters described by the microphysical model and those described by the common core theory in terms of an effective energy scale and an effective mass, if there is any need to do so.

Given that the physics of the problem dictates that all of these various field theory proposals can be effectively described with a massive scalar field, there are real pragmatic and epistemic gains that can be made by leveraging this common core model for the simplest versions of dark energy. In contrast with the overarching approach of inflation, here there is a good argument to be made that it is not necessary to continue to model-build or to use specific microphysical models within the quintessence paradigm as the common core theory offers a perspicuous interpretation of quintessence physics in terms of the microphysics of a massive scalar field. 

Does this break the underdetermination problem in dark energy? In one sense, the answer is clearly `no', as this reasoning does not pick out one of the candidate models of quintessence as more empirically or evidentially favored over the others (and indeed as we have discussed this does not seem to be possible given the current data and epistemological access to cosmological phenomena). However, in the sense of Sec.~\ref{Sec:underdetermination}, we can argue that identifying and moving towards the common core presents us with a privileged ontology for dark energy physics on cosmological scales within this particular class of theories. That is, (i) the common core theory is robust in the sense that it represents a unique description of dark energy physics that all of these various dark energy proposals flow into it within the cosmological regime of interest. Even if one particular theory in the class is rendered unviable (for any reason), the common core theory still survives as a viable description beyond the individual merits of each specific member of the class. Thus, we can be more confident that this ontology is explanatorily useful and viable when compared with the other members. Additionally, (ii) the reality that cosmological data cannot empirically differentiate between these theories and that the cosmological data is only sufficiently course-grained to probe the quadratic order in the potential invites us to consider the scale-relative ontology that is actually accessible to us. If any microsphysical theory in this class will be indistinguishable from the massive scalar field, there is nothing lost by adopting this ontology on those scales. That is, similar to adopting the ontology of fluids on macroscopic scales rather than that of the constituent atoms and molecules, there is nothing lost by adopting the ontology of the massive scalar field on cosmological scales over that of some exotic potential which cosmological data is insensitive to anyway. Those fine-grained details simply do not have an empirical/explanatory/modeling role on cosmological scales (again given the quantity and quality of the data) in a similar manner to how microscopic degrees of freedom often get completely washed out on macroscopic scales. And finally, (iii) as discussed the common core theory has significant pragmatic and epistemic upside due to its simplicity and unificatory power.

In sum, the common core theory arguably goes some way towards ameliorating the underdetermination problems in dark energy research upon adopting the philosophical/non-empirical arguments outlined here as well as considering ontology from the scale-relative perspective. On the other hand, if we want to think more broadly, this reasoning clearly only applies locally within this specific sub-space of theories. In other words, we still have to reckon with the permanent underdetermination between dark energy models described by the theory above and all of the other distinct dark energy proposals that do not fall within this remit (such as more exotic scalar field models, modified gravity models, or even more heterodox proposals \cite{Wolf:2024aeu}).\footnote{This, of course, is just to acknowledge Reutsche's point that ``even explicit RG results are only as reassuring as the space of theories on which the RG group acts is comprehensive.'' \parencite[p.\ 1187]{Ruetsche2018-RUERGR}.}

Finally, one may wonder whether the same (or similar) common core approach could be fruitfully applied to inflation. However, this cannot be the case because the physical situations themselves are quite different. The current dark energy driven epoch has only just begun, meaning that if it is driven by a scalar field the scalar field has traversed only a short stretch of its potential. This is precisely why so many distinct models admit of the Taylor expansion considered here; $\Delta \varphi \simeq 0$ and there are infinitely many distinct potentials that will appear to be arbitrarily close to each other over such short stretches of evolution. By contrast, the inflationary epoch has already undergone its full evolution so we must also consider the details of how inflation ended, meaning that an analogous approach won't capture all of physical details we must consider.

\section{Conclusions}\label{Sec:discussion}

In this article, we have considered the underdetermination present in modern day cosmological modelling of both inflation and dark energy. We have identified this in both cases as an instance of permanent underdetermination in the sense of \textcite{Pitts2010-PITPUF}, and have built upon the analysis of \textcite{FWR} by illustrating in detail how the simplest classes of inflation and dark energy models are underdetermined with respect to their primary observables and by situating this problem within the broader underdetermination literature. Furthermore, noting also that both inflation and dark energy modelling can be understood (and, indeed, often \emph{are} understood by practicing cosmologists) via the framework of EFTs, we have exploited this framework in order to explore how certain philosophical responses to underdetermination might be brought to bear on each case.

Our conclusions offer both good and bad news. The good news is that, in the case of dark energy models, the common core strategy can be applied locally to the quintessence paradigm once one notices that the phenomenology of the distinct microphysical models within it is captured by just the first couple of terms in the expansion of the potential $V(\varphi)$---so, there is little (if anything) to be lost in committing to just such terms in one's ongoing physical reasoning---these terms of course constituting the `common core' of the dark energy models under consideration. Similarly, there might be a viable discrimination strategy for inflation if the observational predictions fall within what we expect for Higgs inflation. On the other hand, the more deflationary news is that the `overarching' strategy which is sometimes adopted in response to the permanent underdetermination of inflationary models seems insufficient to constitute a plausible resolution to this underdetermination, since it is little more than the combination of all such inflationary models into one `larger' model in which some parameters are left unfixed.\footnote{Cf.\ \cite{Maudlin1996-MAUOTU-2} on unification. According to Maudlin, we have unification in a merely unphysical sense if the unification combines multiple physical models without giving some physical account of the common origin of the structures involved in those models, physical interactions between them, etc.} While undeniably useful to the practicing cosmologist, this approach is unable to make a substantive dent in the underdetermination issues highlighted here. And finally, the analysis here of course applies only `locally' within the classes of theories considered here, and does not, for example, address how underdetermination might be dealt with when these theories are compared to other approaches to modelling the phenomena that inflation and dark energy are taken to represent.

Stepping back somewhat, in our view this works represents a fruitful interaction between modern cosmology and philosophy of science. On the one hand, cosmology illustrates live and serious cases of underdetermination that can be  leveraged by philosophers in order to better understand scientific methodology as it is applied by practitioners in real time. On the other hand, philosophy can perhaps provide an illuminating perspective on the epistemic value and pursuit-worthiness of certain approaches given the unique epistemic challenges faced by modern cosmology. For example, one conclusion of our work would be that there is little obviously to be gained at the present moment from further detailed dark energy model-building, or utilizing models other than the common core theory, at least at the level of investigating cosmological phenomena within the quintessence paradigm. Another would be that there perhaps \emph{is} more to be gained from model-building in the inflationary cases, especially with regard to e.g.\ non-minimally-coupled Higgs models, in whose favour various arguments (e.g.\ consilience) would certainly speak. 
%And a final conclusion would be that, if it can be done, developing an EFT for inflation more analogous to the common core theory of dark energy might also be a profitable line of inquiry.

\section*{Acknowledgments}

We are grateful to Pedro Ferreira for many valuable discussions on this topic,  and to audiences in Lugano, Oxford, and Utrecht for helpful feedback. We also thank the two anonymous reviewers. W.W.~acknowledges support from Centre for the History and Philosophy of Physics at St.~Cross College, University of Oxford and the British Society for the Philosophy of Science.

\printbibliography
\end{document}